\documentclass[a4paper,prb,showpacs,twocolumn]{revtex4}

\usepackage{amsfonts}
\usepackage{graphicx}
\usepackage{color}
\usepackage{amsmath}
\usepackage{amssymb}
\usepackage{latexsym}
\usepackage{psfrag}
\usepackage{bbold}
\usepackage[normalem]{ulem}

\begin{document}
\title{Edge channels in graphene Fabry-P\'{e}rot interferometer}
\author{S. Ihnatsenka}
\affiliation{Department of Science and Technology, Link\"{o}ping University, SE-60174, Norrk\"{o}ping, Sweden}
\email{sergey.ignatenko@liu.se}

\begin{abstract}
Quantum-mechanical calculations of electron magnetotransport in graphene Fabry-P\'{e}rot interferometers are presented with a focus on the role of spatial structure of edge channels. For an interferometer that is made by removing carbon atoms, which is typically realized in nanolithography experiments, the constrictions are shown to cause strong inter-channel scattering that establishes local equilibrium and makes the electron transport non-adiabatic. Nevertheless, two-terminal conductance reveals a common Aharonov-Bohm oscillation pattern, independent of crystallographic orientation, which is accompanied by single-particle states that sweep through the Fermi energy for the edge channels circulating along the physical boundary of the device. The interferometer constrictions host the localized states that might shorten the device or disrupt the oscillation pattern. For an interferometer that is created by electrostatic confinement, which is typically done in the split-gate experiments, electron transport is shown to be adiabatic if the staggered potential is introduced additionally into the model. Interference visibility decays exponentially with temperature with a weaker dependence at low temperature. 

\end{abstract}
\pacs{72.80.Vp, 73.43.-f, 85.35.Ds, 73.23.-b}
\maketitle

\section{Introduction}

Quantum Hall interferometers that operate on the Aharonov-Bohm effect have recently been demonstrated in graphene, with high visibility and no Coulomb charging effects.\cite{Dep21, Ron21} This suggests graphene-based interferometers as a better platform for studying the exchange statistics of anionic quasi-particles\cite{Nay08} in comparison to the traditional GaAs-based counterpart.\cite{Cam07-GaAs, Wee89} The conductance oscillations that were measured in Refs. \onlinecite{Dep21, Ron21} were well-described by a theoretical model that is based on an assumption of idealized one-dimensional channels circulating along the edges of the device in the quantum Hall effect (QHE) regime.\cite{Hal82, Wee89, Siv89, Hal11} While good agreement between experiments and the theory seemingly validates the chosen model, or at least does not disprove it, the lack of the spatial structure of the edge states and disregard for electron scattering at the constriction regions in the phenomenological modeling leaves an open question about the physical mechanisms behind the electron interference in the studied devices. The problem is evidenced by strong electron scattering that occurs at the graphene interfaces (i.e., the regions where either device size or crystollographic orientation changes) that has been observed in graphene nanoribbons,\cite{Ihn21} constrictions\cite{Mun06, Ter16, Gui12} and other structures.\cite{Wur09, Wak01-Lib16} As was already pointed out in Ref. \onlinecite{Dep21}, ''quantum Hall interferometer experiments require a precise knowledge of the edge-channel configuration''. Therefore, getting this knowledge, particularly due to massless Dirac fermions in the QHE regime, is necessary for both the interpretation of the interferometry experiments and for the foundation of theories such as those in Refs. \onlinecite{Dep21, Ron21} and \onlinecite{Hal11, Wee89, Siv89}.

Previous studies of mesoscopic graphene devices operating in the QHE regime have addressed energy structure, electronic states and transport in nanoribbons,\cite{Bre06-2, Oos10, Pou10, Per06, Gui12} p-n heterojunctions,\cite{You09-Wei17, Jo22} rings,\cite{Wur10} and others.\cite{Wur09, Mil19-Ngu19, Coi22} These studies have evidenced the existence of edge states,\cite{Hal82} which flow in only one direction along the physical edge of the sample. Edge states flowing in an opposite direction exist at the opposite edge, and it is the absence of scattering between these two edges that constitutes the fundamental reason for the robustness of the quantization of QHE.\cite{Datta} In graphene, the relativistic nature of charge carriers manifests in the so-called anomalous QHE with Landau level (LL) present at zero energy, which separates states with hole character from states with electron character.\cite{Nov05, Cas09} The edge states with the same index of propagating mode, following the standard terminology,\cite{Bee91} are referred to in this study as an edge channel.

This manuscript will provide a microscopic theory of edge channel transport in a graphene interferometer operating on the Aharonov-Bohm effect and will also elucidate the role of the spatial structure of the edge states in electron quantum interference. To this end, the tight-binding model of graphene placed in a perpendicular magnetic field is employed for numerical quantum transport calculations. The interferometer's geometry is created from an infinite graphene nanoribbon, either by removing carbon atoms or by electrostatic confinement in such a way that a square central region is formed between two narrow constrictions, similarly to the Fabry-P\'{e}rot device;\cite{Dep21, Ron21} see the inset in Fig. 1(b). These two types of lateral confinement correspond to a fabrication technique based on etching nanolithography\cite{Bis15-Sar21-Han07, Ter16, Oos10} and split-gates.\cite{Dep21, Ron21} For both cases, quantum transport calculations reveal a common Aharonov-Bohm (AB) interference pattern\cite{Dep21, Ron21, Siv89, Wee89, Bee91, Ihn08, Cam07-GaAs, Datta, Jo22} in conductance, which is due to edge channels circulating along device physical boundaries and scattered at the constrictions. Every conductance peak corresponds to the single-particle state sweeping through the Fermi energy. Conductance oscillations are independent of the crystallographic orientation of the graphene lattice. In contrast to traditional GaAs-based devices, where electron transport in the QHE regime is adiabatic,\cite{Datta, Bee91, Cam07-GaAs, Ihn08, Hal11, Wee89, Hal82, Siv89} AB interference in graphene interferometers that is made by removing carbon atoms occurs because the edge states propagate non-adiabatically and equilibrate locally at the constrictions (interface regions). Relatively strong confinement for Fermi electron gas in graphene is found to cause electron localization along the constriction and might cause a short circuit or deviation in the AB interference signal. In the case of electrostatic confinement, for a device of the same geometry and the staggered potential model, transport is adiabatic, similar to that found in traditional GaAs-based interferometers.\cite{Cam07-GaAs, Ihn08, Hal11, Wee89} The partial penetration of the electron wave function into the potential barriers can result in out-of-phase oscillations of the edge channels. Interference visibility decays overall exponentially with $T$, with weaker dependence at low $T$, in agreement with recent experiments.\cite{Dep21, Jo22}

This manuscript is organized as follows. The theoretical model is formulated in Sec. \ref{sec:model}. The results are presented together with their interpretation and implications for experiment in Sec. \ref{sec:results}. The main conclusions are summarized in Sec. \ref{sec:conclusion}.

\section{Model} \label{sec:model}

The model is based on the standard nearest-neighbor tight-binding Hamiltonian on a honeycomb lattice
\begin{equation}
H = \sum_i \epsilon_i  a_i^{\dagger} a_i -\sum_{\langle i,j \rangle} t_{ij} (a_i^{\dagger}a_j + \textrm{H.c.})
\label{eq:H}
\end{equation}
where $\epsilon_i$ is the on-site energy, $a_i^{\dagger}$ ($a_i$) is the creation (destruction) operator of the electron on the site $i$ and the angle brackets denote the nearest neighbour indices. The magnetic field, $B$, is included via Peierls substitution 
\begin{equation}
t_{ij}=t\exp(i\frac{2\pi}{\Phi_0}\int_{\mathbf{r}_i}^{\mathbf{r}_j}\mathbf{A}\cdot d\mathbf{r}),
\label{eq:tij}
\end{equation}
where $\mathbf{A}=B(-y,0,0)$ is the vector potential in the Landau gauge, $\mathbf{r}_i$ is the coordinate of the site $i$, $\Phi_0=h/|e|$ is the flux quantum, $t=2.7$ eV. Hamiltonian \eqref{eq:H} with $\epsilon_i=0$ is known to describe the $\pi$-band dispersion of graphene well at low energies,\cite{Rei02} and has been used in numerous studies of electron transport in graphene nanostructures.\cite{Cas09, book, Wur09, Wak01-Lib16, Pou10, Wur10} 

Effects due to the next-nearest neighbor hopping, spin, electron-electron interactions are outside of the scope of this study.

The Green's function of the system connected at its two ends to the semi-infinite leads is written as\cite{Datta}
\begin{equation}
\mathcal{G}(\epsilon)=[I\epsilon - H - \Sigma_L(\epsilon) - \Sigma_R(\epsilon)]^{-1}.
\label{eq:Green}
\end{equation}
Here, $H$ describes the scattering region that includes the interferometer itself and a part of the leads, $I$ is the unitary operator, $\Sigma_L(\epsilon)$ is the self-energy due to the semi-infinite left lead at electron energy $\epsilon$, and $\Sigma_R(\epsilon)$ is similarly for the right lead. The lead self-energies are obtained from the surface Green's functions by the method given in Ref. \onlinecite{Xu08}. The system is supposed to be whole graphene made, including the leads. 

Having $\mathcal{G}(\epsilon)$ calculated allows one to obtain observable quantities, like density of states (DOS) and conductance.\cite{Datta} The local density of states (local DOS) for the $i$-th site is given by the diagonal elements of the Green's function as
\begin{equation}
\rho_i(\epsilon)= -\frac{1}{\pi} \textrm{Im}[\mathcal{G}_{ii}(\epsilon)].
\label{eq:LDOS}
\end{equation}
The two-terminal (Hall) conductance $G$ of the system is obtained from Landauer-B\"{u}ttiker formula, which relates conductance to the scattering properties of the system\cite{LB}
\begin{equation}
G=\frac{2e^2}{h}\sum_{\beta\alpha} t_{\beta\alpha}=\frac{2e^2}{h}\sum_{\beta\alpha} \frac{v_{\beta}}{v_{\alpha}} |s_{\beta\alpha}|^2,
\label{eq:G}
\end{equation}
where $t_{\beta\alpha}$ is the transmission coefficient from incoming state $\alpha$ in the left lead to outgoing state $\beta$ in the right lead, $s_{\beta\alpha}$ is the corresponding scattering amplitude, $v_{\alpha}$ and $v_{\beta}$ are the  group velocities for those states, all at the Fermi energy $\epsilon_F$. $s_{\beta\alpha}$ is obtained from the Green's function that connects the first and last slices of the scattering region, see Appendix \ref{app:B}. Another quantity of interest is the probability of electron density $|\Psi_{\alpha}|^2$ (the wave functions modulus), which is obtained from the wave functions in the leads, $s_{\beta\alpha}$ and the Green's function \eqref{eq:Green} by applying Dyson equation as described in Appendix \ref{app:B}.

\section{Results} \label{sec:results}

The system studied is a graphene interferometer that is made from a nanoribbon in an armchair or zigzag configuration by trimming (etching) carbon atoms away or by applying electrostatic potential, see the inset in Fig. \ref{fig:1}. Two (identical) constrictions define the central region similarly to an open quantum dot. For simplicity, the results are presented for rectangular shaped constrictions; a smooth constriction will be commented on. Adopting zigzag and armchair terminology from underlying nanoribbon structure, the interferometer is below referred to as either a zigzag or armchair. The operation regime is chosen to support three channels for electron propagation within which electron can interfere. For 50 nm wide ribbon, which serves as an electron reservoir for the channels, this is achieved at $\epsilon_F=0.2$ $t$ and $B=155..180$ T.\cite{largeB}  Appendix \ref{app:A} elaborates on the propagating states in the chosen regime. The temperature is $T=0$ K unless otherwise stated.

\subsection{Edge channel interference}

\begin{figure}[ht]
\includegraphics[keepaspectratio,width=\columnwidth]{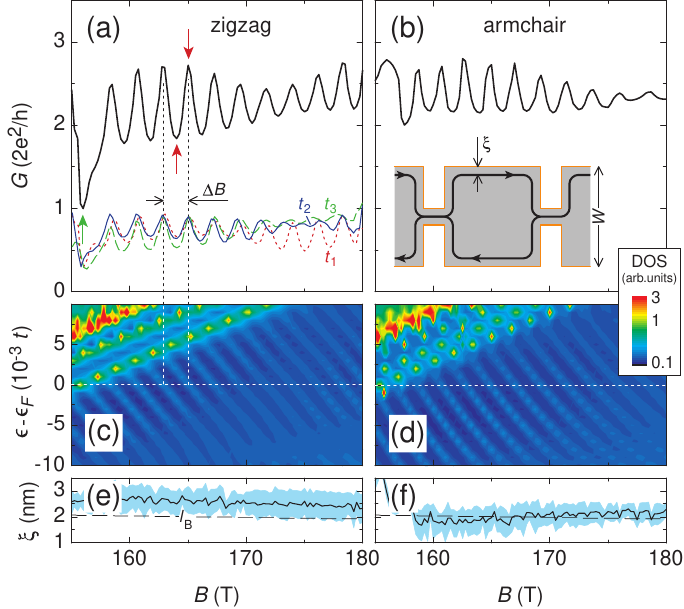} 
\caption{Conductance $G$, single-particle state spectroscopy and edge channel displacement $\xi$ in graphene interferometers with zigzag (left-hand panels) and armchair (right-hand panels) orientation. (a,b) $G$ oscillates as a function of magnetic field $B$ with peaks matching the crossings of the resonant energy levels and the Fermi energy $\epsilon_F$ in (c,d). The energy levels can be traced by enhanced DOS in (c,d), which is obtained by integrating local DOS over the central region of interferometer (a dot in between of two constrictions). Another set of the energy levels, with the positive slope, is due to the states localized at the constrictions. The geometrical area of the dot, reduced by $\xi$, whose evolution is plotted in (e,f), defines an interfering area that is enclosed by the clockwise propagating edge channel as illustrated in the inset between (a) and (b). The dot geometry is a square with sides $W=50$ nm. The arrows in (a) mark $B$ for which the edge states are shown in Figs. \ref{fig:2} and \ref{fig:6}. The dashed line in (e,f) is the magnetic length $l_B$; blue filled area denotes the standard deviation.}
\label{fig:1}
\end{figure}

Numerical calculation of quantum electron transport, described by Eqs. \eqref{eq:H}-\eqref{eq:G}, reveals conventional conductance oscillations\cite{Ihn08, Bee91, Datta, Wee89, Cam07-GaAs} in a graphene AB interferometer, irrespective of crystallographic orientation; see Figs. \ref{fig:1}(a),(b). The peaks in $G$ correspond to the resonant states passing through $\epsilon_F$, similar to what was found in Refs. \onlinecite{Wee89, Ihn08}. These resonant states can be traced in Figs. \ref{fig:1}(c),(d) as bright trenches with a negative slope. Sloping downward with increasing $B$ implies that degeneracy of the occupied LLs increases via their edge states, provided by the geometry confinement that makes LLs to rise in energy on approaching the sample boundary.\cite{Bee91, Datta} 
Downward sloping corroborates AB regime of interference, as opposed to Coulomb dominated regime.\cite{Cam07-GaAs, Hal11} For each LL, the degree of degeneracy is quantified by the number of states per unit area, which increases as $B/\Phi_0$. Every resonant state is a result of the constructive interference of the electron wave in the edge channels that are backscattered at the two constrictions.
The difference between phases in two arms of the interferometer is proportional to the total flux $\Phi$ enclosed by the area $S$ of the interfering path;\cite{Datta} $\Phi=BS$. Changing $\Phi$ by $\Phi_0$, via applied $B$, causes the phase difference to accumulate a value of $2\pi$ and $G$ to develop one oscillation period.\cite{Datta, Wee89}
One period, $\Delta B=2.22$ and 2.16 T for zigzag and armchair configuration, yeilds area, $S=1872$ and $1924$ nm$^2$, that is less than the geometrical area of the central region 2500 nm$^2$. This discrepancy might be attributed to a finite spatial extent of the edge channel, so the interfering area is smaller then the geometrical one. Figs. \ref{fig:1}(e),(f) substantiate this argument by showing an averaged edge channel distance from the physical boundary $\xi=[\langle\mathbf{r}|\psi_{\alpha}\rangle]_{\alpha,x}$, where averaging is done over channels and the straight segments along the boundaries. $\xi$ is about the magnetic length $\l_B$ and, interestingly, doesn't reveal any clear beat of $\Delta B$ as it was argued to occur in the Coulomb dominated regime.\cite{Hal11} Slightly larger $\xi$ for zigzag orientation, in comparison to armchair one, explains slightly smaller $S$ and larger $\Delta B$. Transmission coefficients (dimensionless) for individual edge channels $t_{\alpha}$, as shown in Fig. \ref{fig:1}(a), reveal in-phase oscillations of nearly equal amplitude. Therefore, $G$ oscillations in graphene AB interferometer are due to simultaneous interference of the edge channels propagating at about $\l_B$ distance from the device boundaries.

Because zigzag and armchair interferometers reveal qualitatively similar dependencies for $G$ and the structure of energy levels, below only zigzag configuration is considered.

\begin{figure}[t]
\includegraphics[keepaspectratio,width=\columnwidth]{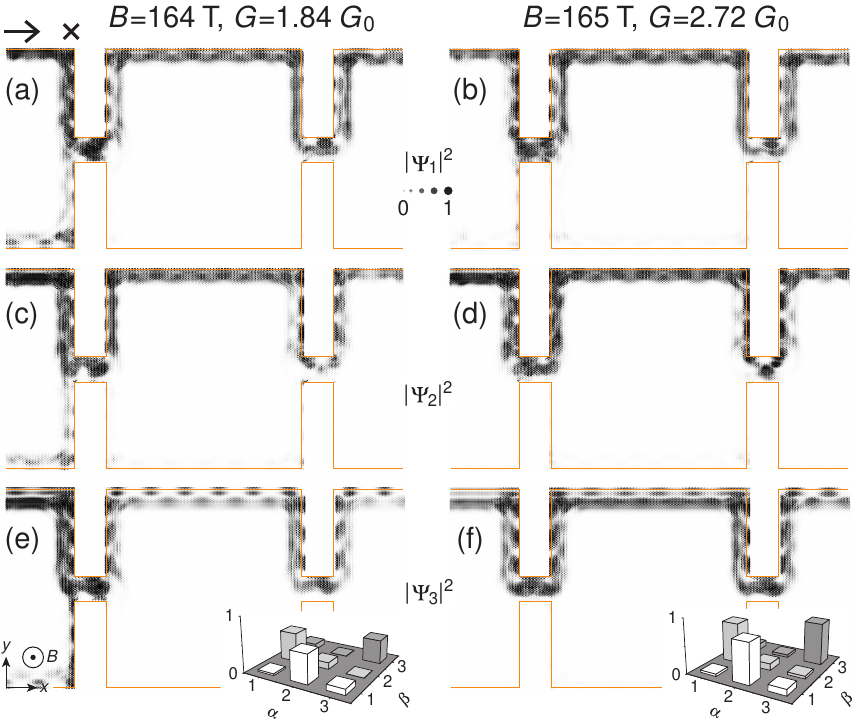} 
\caption{The wave function modulus $|\Psi_{\alpha}|^2$ of $\alpha$-th incoming state for $B$ marked by the red arrows in Fig. \ref{fig:1}(a). The left-hand (right-hand) panels represent the minima (maxima) of $G$ oscillation. In (a), the arrow and cross illustrate schematically the incoming state and forward scattering to another states that occurs at the constriction bend. The insets in (c,f) show the transmission coefficients from incoming $\alpha$ to outgoing $\beta$ state. $G_0=2e^2/h$.}
\label{fig:2}
\end{figure}

Figure \ref{fig:2} shows the edge states characterizing conductance oscillation at its peak and dip values.
The details on the electronic states entering and leaving the interferometer, which are at the openings in these plots, are given in the Appendix \ref{app:A}. Edge channel visualization leads to immediate conclusion: A distinct feature of electron transport in a graphene interferometer in comparison to a conventional GaAs-based interferometer\cite{Cam07-GaAs, Wee89, Ihn08, Hal11} is \textit{non-adiabaticity due to strong scattering between the edge states that occurs at the constrictions}. Even though $G$ varies by nearly one conductance quanta over $B$ interval in Figs. \ref{fig:1}(a),(b), the oscillations are not caused by the highest occupied LL edge state (as it does for a conventional GaAs-based interferometer), but rather by a mix of all of the states, corroborated by $t_{\alpha}$ in Fig. \ref{fig:1}(a). Note that all plots on the left panel has visible reflected wave when compared to the corresponding right panel plots in Fig. \ref{fig:2}. Strong inter-channel scattering is further evidenced by the transmission coefficients in the insets in Figs. \ref{fig:2}(c),(d). This indicates local equilibrium\cite{Bee91} of the edge states due to graphene interfaces. The effect develops clearly at the constriction bending as illustrated by the arrow and cross in Fig. \ref{fig:2}(a). It is essentially the same for all bends along the edge channel path.
If rectangular shaped constriction is smoothed out toward a cosine-like profile the results do not change qualitatively thus implying that non-adiabaticity is a result of abrupt lattice termination.

Another observation in Fig. \ref{fig:2} is valley selective scattering ($t_{2\leftarrow 1}>t_{3\leftarrow 1}$) in the graphene interferometer, which is in line with conclusion about valley degeneracy lifting of AB interference in graphene rings.\cite{Wur10}

\begin{figure}[t]
\includegraphics[keepaspectratio,width=0.95\columnwidth]{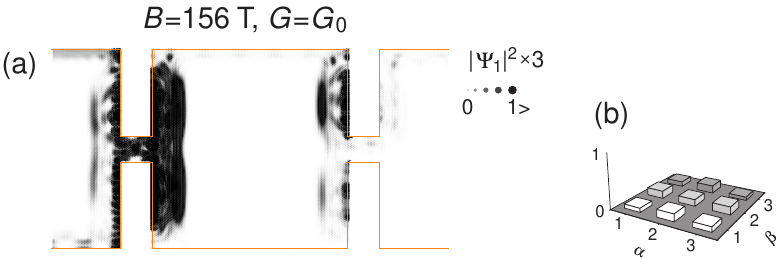} 
\caption{Short circuit of the interferometer by the localized states at the constrictions. (a) The first propagating state $\alpha=1$ with $|\Psi_1|^2$ magnified and truncated (for the sake of better visualization). Similarly strong (exponential) localization at the constriction occurs for all other propagating states; the transmission coefficients are shown in (b). $B=156$ T is marked by the green arrow in Fig. \ref{fig:1}(a).}
\label{fig:6}
\end{figure}

The interferometer constrictions, acting as scattering centers for incident electrons, host another set of the single-particle states that rather depopulate in increasing $B$ --- visible as bright trenches with a positive slope in Figs. \ref{fig:1}(c)(d). These states are localized along the constrictions and cause shorting in case of the zigzag interferometer due to their coupling to and backscattering the incident states in the edge channels at the entrance constriction: $G$ drops to $2e^2/h$ at $B=156$ T in Fig. \ref{fig:1}(a), see Fig. \ref{fig:6}. For the armchair interferometer, AB oscillation periodicity is seemingly violated at $B$, for which the localized states cross $\epsilon_F$. The difference between armchair and zigzag interferometers might be attributed to the different atomic arrangements of the edges, and consequently to different low-energy electronic states causing localization.\cite{Bre06} Note that the armchair interferometer in the present study is crystallographically an inverse of a zigzag one in the sense that all of the zigzag edges are replaced by armchair edges, and vice versa. It is known that depending in the coupling details between the electronic states transmission through a mesoscopic system can reveal resonances or anti-resonances.\cite{Emb99} The existence of the localized states at the constrictions and the shorting in case of the zigzag interferometer, Fig. \ref{fig:6}, imply that the strong electron backscattering on the graphene interfaces, which has been previously observed for different structures at zero $B$,\cite{Wur09, Wak01-Lib16, Ihn21} also persists in the QHE regime.

\subsection{Electrostatic confinement}

The Fabry-P\'{e}rot interferometers that were studied experimentally in Refs. \onlinecite{Dep21, Ron21} were made from a uniform graphene layer rather by imposing electrostatic confinement. In those studies, the fabricated devices contained additional split gate electrodes that expelled the charge carriers from the area beneath by shifting $\epsilon_F$ into the energy gap between LLs. The gap formation was controlled in separate measurements to occur at the graphene charge neutrality point in the magnetic field, arguably due to electron interactions. The electron interactions are known to cause splitting of 4-fold degenerated zeroth LL,\cite{Zha09, Gus09, Zim16} whose degeneracy comes from electron spin and graphene valley, such that the insulating state develops throughout interior of graphene layer. To address this regime theoretically, the simplest approach is to brake inversion symmetry of graphene lattice by adding the staggered potential to the on-site term $\epsilon_i$ in the Hamiltonian \eqref{eq:H}: A positive value $\Delta$ is added on one atom of the graphene unit cell and negative on the other. A value $\Delta=0.04t$ is chosen to make a sufficiently opaque tunnelling barrier due to split gates and the interfering path to be well defined for the device sizes studied here. Instead of removing carbon atoms  an electrostatic potential $V=\epsilon_F$ is applied to those atoms, and the constrictions are made slightly narrower to couple the counter-propagating edge channels. Other than that the interferometer geometry and parameters are the same as in the previous section. 

\begin{figure}[t]
\includegraphics[keepaspectratio,width=0.65\columnwidth]{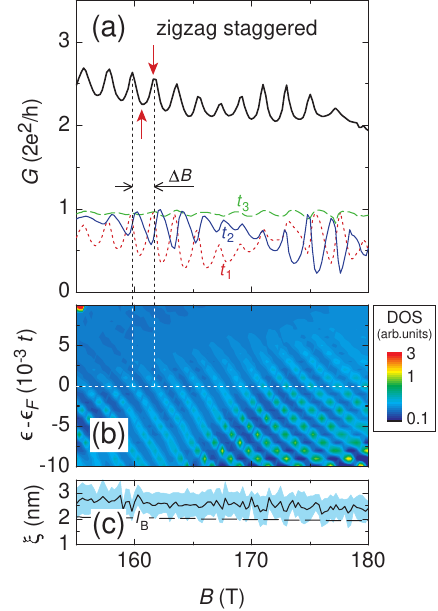} 
\caption{The same as (a),(c),(e) in Fig. \ref{fig:1} but for the electrostatic confinement with the staggered potential.}
\label{fig:81}
\end{figure}

\begin{figure}[t]
\includegraphics[keepaspectratio,width=\columnwidth]{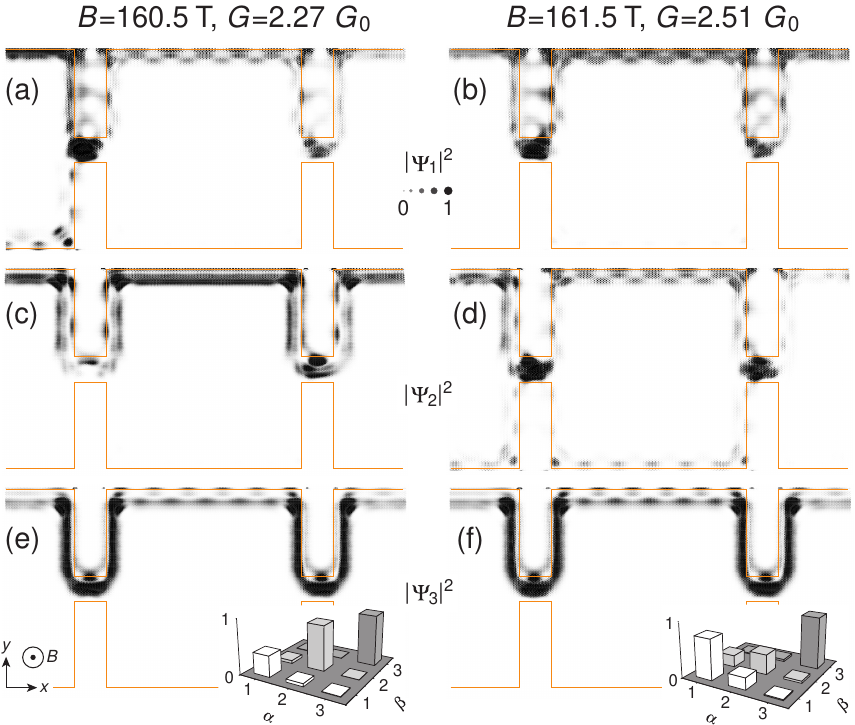} 
\caption{The same as Fig. \ref{fig:2} but for the electrostatic confinement and $B$ marked by the red arrows in Fig. \ref{fig:81}(a).}
\label{fig:33}
\end{figure}

Figure \ref{fig:81} shows conductance, single-particle state spectroscopy and edge channel displacement in zigzag graphene interferometer with electrostatic confinement. Compared to the etched devices, several features of electrostatic confinement can be observed. The first is the smaller oscillation period even though the system sizes remain the same: $\Delta B=1.95$ T compared to 2.22 T in Fig. \ref{fig:1}(a). This implies larger interfering area $S=2122$ nm$^{-2}$ (compared to 1872 nm$^{-2}$). Visualization of the wave function in Fig. \ref{fig:33} reveals that the interfering path becomes smooth and partially penetrates into the potential barriers of the constrictions, cutting off their corners. The first edge channel, $|\Psi_1|^2$, while developing a loop, leaks also through the gated areas, Fig. \ref{fig:33}(a),(b). The second feature is \textit{the adiabatic transport due to weaker confinement strength}, for which the transmission coefficients of the individual channels $t_{\alpha}$ show little inter-channel scattering, see the insets in Figs. \ref{fig:33}(c),(f). Thirdly, $t_{\alpha}$ oscillate out-of-phase for some $B$, c.f. $\alpha=1$ and $\alpha=2$ in Figs. \ref{fig:81}(a) and \ref{fig:33}(a)-(d). This might be attributed to a  partial leakage through the gated areas, when tunneling occurs not exactly at the narrowest point in the constriction. As a result, $G$, being a sum over all $\alpha$, reveals an oscillation amplitude that can be smaller than that of individual edge channels. As the device sizes increase, the edge channels are expected to follow gate boundaries precisely, $t_{\alpha}$ to oscillate in-phase and $G$ amplitude to increase in the regime of multiple edge channels, similar to the adiabatic transport regime in GaAs-based interferometers.\cite{Datta, Bee91, Cam07-GaAs, Ihn08, Hal11, Wee89, Hal82, Siv89} This is likely a regime realized in the experiments in Refs. \onlinecite{Dep21, Ron21}. 

\subsection{Visibility}

\begin{figure}[th]
\includegraphics[keepaspectratio,width=0.7\columnwidth]{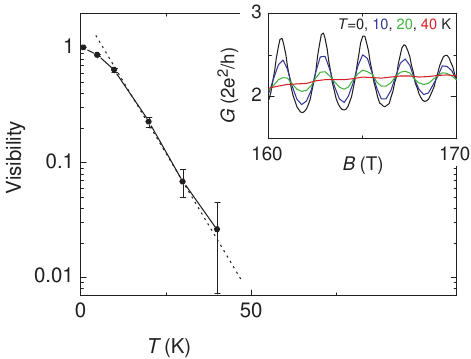} 
\caption{Temperature dependence of the normalized visibility for zigzag interferometer. The error bars correspond to the standard deviation. The straight dotted line shows the exponential dependence and is a guide for eye.
The inset shows smearing $G$ oscillations for several $T$.}
\label{fig:7}
\end{figure}

Figure \ref{fig:7} shows temperature dependence of the interference visibility, normalized by $\nu(T=0)$,
\begin{equation}
\nu=\frac{G_{\textrm{max}}-G_{\textrm{min}}}{G_{\textrm{max}}+G_{\textrm{min}}},
\label{eq:visibility}
\end{equation}
where the effect of $T$ is introduced via the derivative over the Fermi-Dirac function\cite{Datta}
\begin{equation}
G=-\frac{2e^2}{h}\int d\epsilon \frac{\partial f}{\partial\epsilon}G(\epsilon).
\label{eq:GT}
\end{equation}
As $T$ increases, more neighbor resonance states, Fig. \ref{fig:1}(c),(d), contribute to conduction that averages out oscillating amplitude, see inset to Fig. \ref{fig:7}. $\nu$ decays exponentially over two decades, while there is a visible saturation at low $T$, in agreement with recent experimental findings.\cite{Dep21, Jo22}

\section{Conclusion} \label{sec:conclusion}

A quantum-mechanical model of electron magnetotransport in a graphene Fabry-P\'{e}rot interferometer, that explicitly accounts for the spatial structure of electron states and their interference and does not rely on any phenomenological parameters (like transmission amplitudes of the constrictions\cite{Dep21, Ron21, Hal11, Wee89, Siv89}), is presented. For interferometers of different crystallographic orientations, numerical calculations reveal a common Aharonov-Bohm interference effect, irrespective of the orientation of the graphene lattice. Two-terminal conductance oscillates as one magnetic flux quanta is added to the interfering path, accompanied by one single-particle state added to edge channels circulating along the physical boundary of the device. In the case of geometry made by etching, the interferometer constrictions cause strong inter-edge-channel scattering that causes the system to establish a local equilibrium and electron transport to be non-adiabatic. The interferometer constrictions host the localized states that might shorten the device or disrupt the oscillation pattern. Transport, however, is adiabatic in the case of electrostatic confinement, when valley splitting is introduced via the staggered potential. It is similar to traditional GaAs-based interferometers,\cite{Wee89, Cam07-GaAs, Ihn08, Hal11} though the partial penetration of the electron wave function into the potential barriers affects edge channel propagation. Interference visibility decays exponentially with $T$ showing a weaker dependence at low $T$. 

The results suggest that any graphene interface that exposes the physical edges of the graphene lattice acts as an "ideal" contact in the QHE regime. By establishing a local equilibrium, and thus making the edge channels equally populated, a prerequisite is fulfilled for the use of a local resistivity tensor.\cite{Bee91} However, common theories of QHE, where transport is assumed to be adiabatic, are not applicable to such systems or (in particular) to graphene interferometers made by etching nanolithography.

In the experiments in Refs. \onlinecite{Dep21, Ron21}, electrostatic confinement was realized by adjusting depletion regions 
into the energy gap. To account for this beyond the simple model of the staggered potential, many-electron and (possibly) spin\cite{Zha09-Epp19} effects should be added to the theory. Also, for a quantitative comparison, the realistic shape (and size) of the split gates and potential due to those split-gates might be needed. The study presented here might serve as a basis the further development of the graphene AB interferometer theories. 

\section{Acknowledgement}
This work was supported by SNIC 2021/22-961. I thank to B. Sac\'{e}p\'{e} for pointing out their recent work and C. Beenakker for comment on terminology.

\appendix
\section{Edge channels in the leads} \label{app:A}
\begin{figure*}[t]
\includegraphics[keepaspectratio,width=0.9\textwidth]{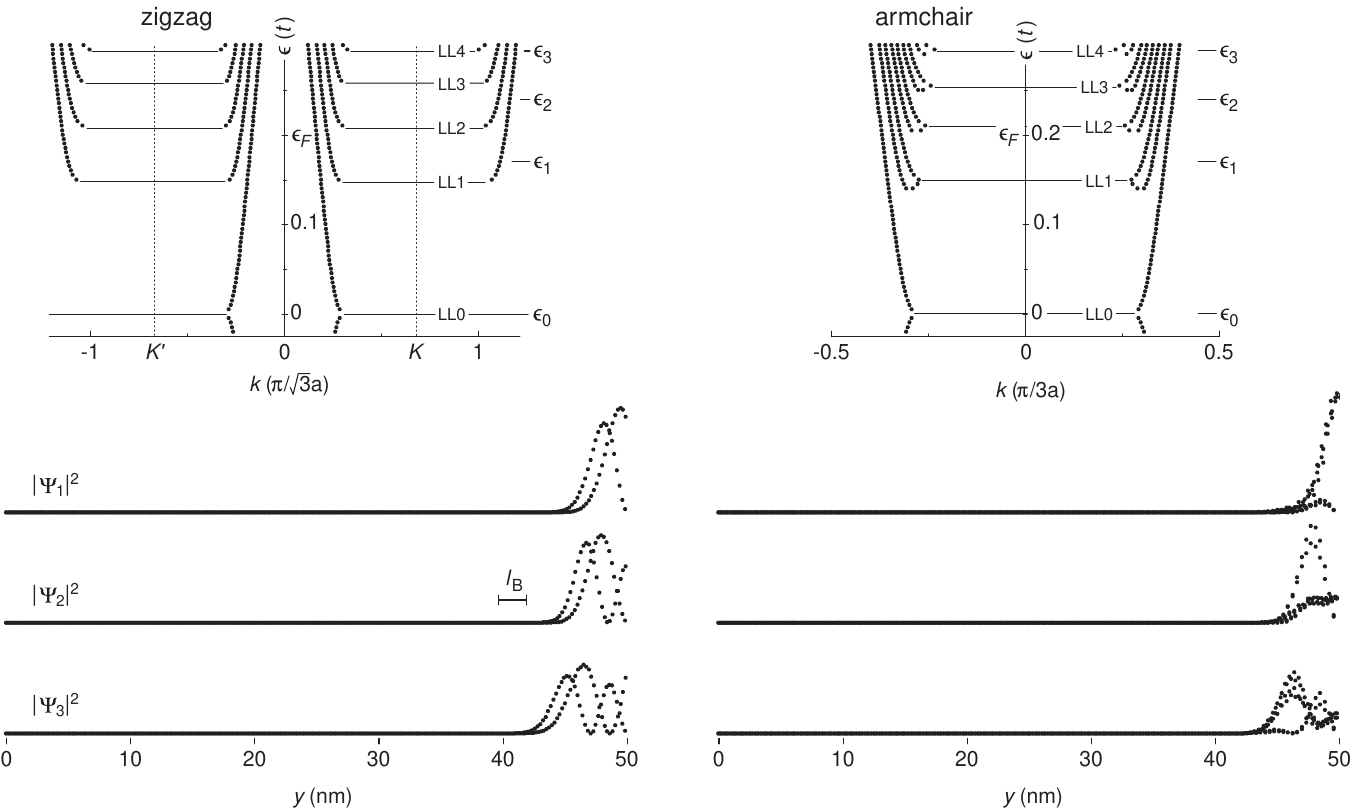} 
\caption{The band structures (top panels) and wave functions (bottom panels) for zigzag and armchair graphene nanoribbons in the QHE regime, $B=160$ T. In the top panels, the horizontal solid lines denote LLs and are guides for the eye; the vertical dotted lines denote $K$ and $K'$ points of the first Brillouin zone of graphene. $K$ and $K'$ points coincide at $k=0$ for the armchair ribbon. $\epsilon_n$ labels $n$-th LL in 2D graphene obtained from the Dirac equation. The bottom panels show the wave function modulus $|\Psi_{\alpha}|^2$ of $\alpha$-th propagating state (projected onto the ribbon cross-section) at $\epsilon_F=0.2$ $t$; $\epsilon_F$ is marked on the corresponding top panels. At $B=160$ T, $l_B=2.03$ nm. $a=0.142$ nm.}
\label{fig:A}
\end{figure*}

Figure \ref{fig:A} shows the dispersion relation and wave functions in graphene nanoribbons --- the structures, which serve as electron reservoirs (leads) in the interferometers studied in the main text. At $B=160$ T,\cite{largeB} the dispersionless bands can be traced, centered around graphene $K$ and $K^{\prime}$ points, and be attributed to Landau levels (LLs).\cite{Bre06-2, Cas09, Per06}  The identified LLs agree reasonably well with the solution of the Dirac equation in 2D graphene\cite{Cas09}
\begin{equation}
\epsilon_n=v_F\sqrt{2e\hbar Bn},
\label{eq:EDirac}
\end{equation}
where $v_F=10^6$ m/s and $n$ is the index of electron LL, counted from $n=0$ at $\epsilon=0$. At $\epsilon_F=0.2$ $t$, zeroth and first LLs are occupied in the ribbon bulk and provide, accounting for the valley degeneracy,\cite{Cas09} three channels for electron propagation along the ribbon edges (in both directions), see also the left hand panels of Figs. \ref{fig:2} and \ref{fig:33}, where the electron states enter the interferometer. In Fig. \ref{fig:A}, the wave functions propagating only in one (positive $x$) direction are shown. (As it is intrinsic for QHE, similar states on the opposite edge propagate in the negative $x$ direction.\cite{Datta, Bee91, Hal82}) The dispersion relations in Fig. \ref{fig:A} reveal electron bands that raise in energy and cross $\epsilon_F$ at momentum whose conversion to coordinate space $y=-hk_x/eB$ gives approximate location of the edge channel.\cite{Datta} The latter thus can be viewed as real space realisation of the Fermi surface.

Recently, the electron edge states and LL spectroscopy have been directly observed in scanning tunnel microscopy measurements.\cite{Coi22} On approaching the sample boundary, LLs were found to raise in energy and edge channels to be sharply confined within few $l_B$ to the physical boundary of graphene, all consistent with Figure \ref{fig:A}.

\section{Scattering problem for graphene interferometer} \label{app:B}
\label{app:B}
In this appendix, transmission coefficients, entering the Landauer-B\"{u}ttiker formula\cite{LB} for conductance \eqref{eq:G}, are derived using the Green's functions. The derivation follows Ref. \onlinecite{Xu08}, and the reader is referred to this reference for further details. Similar approaches can be found in Refs. \onlinecite{And91, Ana08, Rot03, Mac85, Sol89}. The method is essentially the same as one used for the materials with parabolic band dispersion, where the tight-binding model is formulated on the square lattice.\cite{Ihn06}

\subsection{Bloch states in ideal GNR}

\begin{figure}[t]
\includegraphics[keepaspectratio,width=\columnwidth]{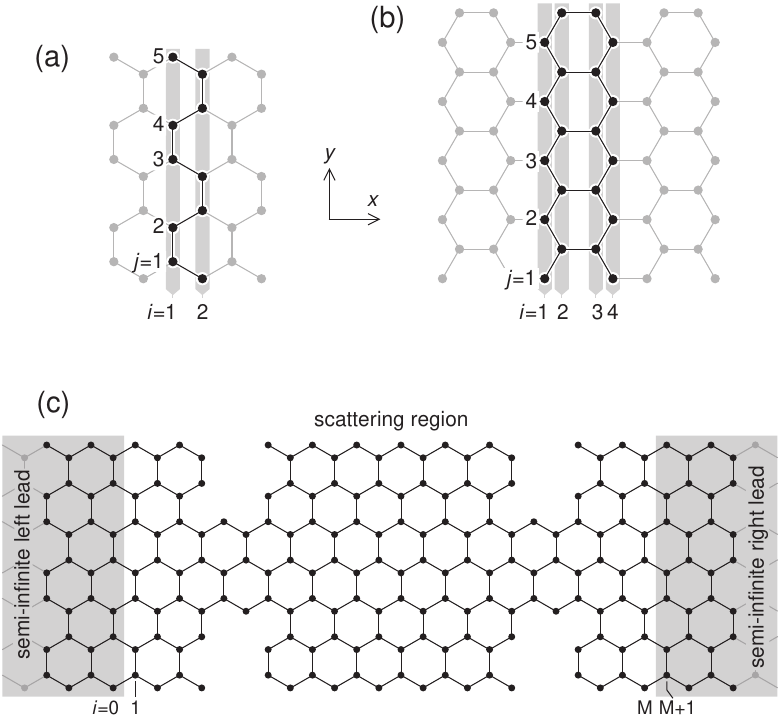} 
\caption{The unit cells of zigzag (a) and armchair (b) GNRs. The unit cell is periodic in the $x$ direction and is chopped into $M$ transverse slices (denoted by the grey shaded areas) in the $y$ direction. $M=2$ and $M=4$ for zigzag and armchair unit cells. In (a) and (b), every slice contains $N=5$ sites. GNR becomes the semi-infinite leads in the computation domain (c), which also includes the scattering region in between. The scattering region, whose slices run $1\leq i\leq M$, can generally be of arbitrary geometry.\cite{Sol89} In present study, it has two constrictions, defined either by removing lattice sites as shown in (c) or imposing potential on those sites.}
\label{fig:B}
\end{figure}

First, let us consider an ideal GNR infinitely long in $x$ direction and consisting of $N$ lattice sites in $y$ direction, Fig. \ref{fig:B}. For graphene hexagonal lattice, two orientations are considered as basic, with terminations along ribbon edges being either zigzag or armchair.\cite{Cas09} The unit cells for these terminations consist of $M=2$ and 4 slices, Fig. \ref{fig:B}(a),(b). Every lattice site corresponds to a carbon atom and, more specially, to its $p_z$ orbital.\cite{Rei02} In this approach, no discrimination applies on A and B graphene sublattices,\cite{Cas09} and whether the orientation is zigzag or armchair depends solely on connection between the lattice sites.

The solution of the Schr\"{o}dinger equation with Hamiltonian \eqref{eq:H} for the ideal GNR can be written in terms of an anzats for a Bloch wave
\begin{equation}
|\psi\rangle=\sum_{\alpha} e^{ik_{\alpha}x}|\chi_{\alpha}\rangle, 
\label{eq:Bloch}
\end{equation}
where $k_{\alpha}$ is the Bloch wave vector in the direction of translation invariance and $|\chi_{\alpha}\rangle$ is the periodic eigenfunction. The sum in \eqref{eq:Bloch} runs over propagating and evanescent states to form a complete set.\cite{LanIII}

For the ideal GNR, the Hamiltonian \eqref{eq:H} can be rewritten as a sum of the operators describing the unit cell, the outside region, and coupling between them
\begin{equation}
H=H_{\text{cell}}+H_{\text{out}}+U.
\label{eq:H3}
\end{equation}
The unit cell term includes the slices $1\le i \le M$, as shown in Figs. \ref{fig:B}(a) and (b), while the term for the outside region goes all over the other slices, $-\infty < i \le 0$ and $M+1 \le i < \infty$. These two terms are coupled by the transfer integrals \eqref{eq:tij} acting between slices $0\leftrightarrow 1$ and $M \leftrightarrow M+1$. For the Hamiltonian written in the form \eqref{eq:H3}, the corresponding wave function is
\begin{equation}
|\psi\rangle = |\psi_{\text{cell}}\rangle + |\psi_{\text{out}}\rangle.
\label{eq:wf3}
\end{equation}
Defining the Green's function in a way suitable for calculations in matrix form\cite{Datta, And91}
\begin{equation}
\mathcal{G}=(E-H+i\eta)^{-1},
\label{eq:Gdefinition}
\end{equation}
with $\eta\rightarrow0^+$, the wave function of the cell can be written as 
\begin{equation}
|\psi_{\text{cell}}\rangle =\mathcal{G}_{\text{cell}}U |\psi_{\text{out}}\rangle,
\label{eq:wf4}
\end{equation}
where $\mathcal{G}_{\text{cell}}$ is the Green's function of the operator $H_{\text{cell}}$. Taking the matrix elements of the wave functions for the first $i=1$ and the last $i=M$ slices of the unit cell, this equation can be written in the matrix form
 \begin{align}
    \psi_1 &=\mathcal{G}_{1,1}U_{1,0}\psi_0 + \mathcal{G}_{1,M}U_{M,M+1}\psi_{M+1}, \label{eq:psi1} \\
    \psi_M &=\mathcal{G}_{M,1}U_{1,0}\psi_0 + \mathcal{G}_{M,M}U_{M,M+1}\psi_{M+1}, \label{eq:psiM}
\end{align}
where $\psi_i$ is the vector column describing the wave function for the slice $i$, $\mathcal{G}_{i,i^{\prime}}$ is the Green's function matrix connecting slices $i$ and $i^{\prime}$, and $U_{i,i^{\prime}}$ is the matrix of hopping integrals \eqref{eq:tij}. Translation invariance of the unit cell implies $U_{M,M+1}=U_{0,1}$, Figs. \ref{fig:B}(a),(b). The Bloch’s theorem in terms of the periodic eigenstates 
\begin{equation}
\chi_{i+M}=e^{ikM} \chi_{i},
\end{equation}
allows further to rewrite Eqs. \eqref{eq:psi1} and \eqref{eq:psiM} as the eigenvalue problem
\begin{equation}
\begin{pmatrix}
-\mathcal{G}_{1,M}U _{0,1}& 0 \\
-\mathcal{G}_{M,M}U_{0,1} & I
\end{pmatrix}^{-1}
\begin{pmatrix}
-\mathbb{1} & -\mathcal{G}_{1,1}U_{1,0} \\
0 & -\mathcal{G}_{M,1}U_{1,0} 
\end{pmatrix}
\begin{pmatrix}
\chi_1 \\
\chi_0 
\end{pmatrix}
= e^{ikM}
\begin{pmatrix}
\chi_1 \\
\chi_0 
\end{pmatrix}
\label{eq:Bloch2}
\end{equation}
where $I$ is the unitary matrix. Eq. \eqref{eq:Bloch2} has $2N$ eigenvalues and $2N$ eigenvectors, which are the Bloch states classified into $N$ right- and $N$ left-going waves. The right-going solutions consists of travelling waves with velocity in positive $x$ direction and evanescent waves decaying exponentially in the positive $x$ direction. Similarly, the left-going solutions consists of propagating and decaying waves in negative $x$ direction.

If the Bloch wave \eqref{eq:Bloch} is a set of the eigenstates of Hamiltonian \eqref{eq:H3}, in which $H_{\text{out}}$ splits into infinite replicas of $H_{\text{cell}}$ connected by $U$, the group velocity for an eigenstate $\alpha$ is
\begin{equation}
v_{\alpha}=\frac{1}{\hbar}\frac{\partial}{\partial k} \langle \psi_{\alpha} | H |\psi_{\alpha}\rangle=\frac{i M}{\hbar} \langle \chi_{\alpha} | Ue^{ikM}-U^{\dag}e^{-ikM} |\chi_{\alpha}\rangle,
\end{equation}
where $U^{\dag}$ is the Hermitian conjugate of the coupling operator, and $|\psi_{\alpha}\rangle$ is the Bloch wave normalized on unit flux.

\subsection{Surface Green's function}
Let's consider a semi-infinite ideal graphene ribbon extending from slice $-m$ to the right, $-m\leq i < \infty$. Suppose that an excitation $|s\rangle$ is applied to its surface slice $i=-m$. Whenever the response $|\psi\rangle$ is related the excitation $|s\rangle$ by a differential operator $D_{\text{op}}$ as $D_{\text{op}}|\psi\rangle=|s\rangle$ we can define a Green's function (propagator) and express the response in the form\cite{Datta}
\begin{equation}
|\psi\rangle=D_{\text{op}}^{-1}|s\rangle=\mathcal{G}|s\rangle,
\label{eq:formalG}
\end{equation}
where $|\psi\rangle$ is the wave function that has to satisfy the Bloch condition \eqref{eq:Bloch}. Consider a unit cell of a graphene lattice, $1\leq i \leq M$, $M=2$ and 4 for the zigzag and armchair orientation, see Fig. \ref{fig:B}(a),(b). Applying the Dyson’s equation between the slices 0 and 1, we obtain
\begin{equation}
\mathcal{G}_{1,-m}=\Gamma_r U_{1,0}\mathcal{G}_{0,-m},
\label{eq:Dyson}
\end{equation}
where $\Gamma_r\equiv \mathcal{G}_{1,1}$ is the right surface Green's function. Evaluating the matrix elements $\langle \psi_1|\psi\rangle$ of \eqref{eq:formalG} and making use of \eqref{eq:Dyson}, we obtain for each Bloch state $\alpha$, $\psi_{1}^{\alpha}=\Gamma_r U_{1,0}\psi_{0}^{\alpha}$. The latter equations can be used for determination of $\Gamma_r$
\begin{equation}
\Gamma_rU_{1,0}=\Psi_1\Psi_0^{-1},
\end{equation}
where $\Psi_1$ and $\Psi_0$ are the square matrixes composed of the column vectors $\chi_1^{\alpha}$ and $\chi_0^{\alpha}$, $1\leq \alpha \leq N$, Eq. \eqref{eq:Bloch2}, i.e. $\Psi_1=(\psi_1^1, ... , \psi_1^N)$, $\Psi_0=(\psi_0^1, ... , \psi_0^N)$. The expression for the left surface Green’s function $\Gamma_l$ (i.e., the surface function of the semi-infinite ribbon open to the left) is derived similarly
\begin{equation}
\Gamma_lU_{1,0}^{\dag}=\Psi_M\Psi_{M+1}^{-1},
\label{eq:Gl}
\end{equation}
where the matrixes $\Psi_M$ and $\Psi_{M+1}$ are defined in a similar way as $\Psi_1$ and $\Psi_0$ above.

\subsection{Transmission coefficients}

To calculate the transmission coefficients and hence evaluate \eqref{eq:G}, the interferometer structure is divided into three regions: two ideal semi-infinite leads of the width $N$ extending in the regions $i\leq0$ and $i\geq M+1$, respectively, and the scattering region, Fig. \ref{fig:B}(c). The latter composes of two constrictions and the cavity in between, see the inset in Fig. \ref{fig:1}(a). In general, the scattering region can contain arbitrary scatterers and be of arbitrary shape.\cite{Datta}

The incoming, transmitted, and reflected states in the leads have the form of Bloch waves \eqref{eq:Bloch}
\begin{align}
|\psi_{\alpha}^\text{i}\rangle &= \sum_{i\leq 0} e^{ik_{\alpha}^+ x_i} |\chi_{\alpha}\rangle \\ \label{eq:s}
|\psi_{\alpha}^\text{s}\rangle &= \sum_{i\geq M+1} \sum_{\beta} t_{\beta \alpha}e^{ik_{\beta}^+ (x_i-x_{M+1})} |\chi_{\beta}\rangle \\ 
|\psi_{\alpha}^\text{r}\rangle &= \sum_{i\leq 0} \sum_{\beta} r_{\beta \alpha}e^{-ik_{\beta}^- x_i} |\chi_{\beta}\rangle,
\label{eq:states}
\end{align}
where $t_{\beta\alpha}$ ($r_{\beta\alpha}$) are the transmission (reflection) amplitude from the state $\alpha$ to the state $\beta$, plus (minus) superscripts for the wavevectors denote right (left) going direction. The sum over $\beta$ includes outgoing propagating and evanescent states.

The solution of the Schr\"{o}dinger equation 
\begin{equation}
H |\psi\rangle=E |\psi\rangle
\end{equation}
with $|\psi\rangle=|\psi_{\alpha}^\text{s}\rangle+|\psi_{\alpha}^\text{i}\rangle$ for the transmitted state in the right lead can be written as
\begin{equation}
|\psi_{\alpha}^\text{s}\rangle=\mathcal{G}(H-E) |\psi_{\alpha}^\text{i}\rangle.
\label{eq:psiH}
\end{equation}
To find the transmission matrix, let us consider the matrix element $\langle\psi_{M+1}|\psi_{\alpha}^{\text{s}}\rangle$. Using \eqref{eq:psiH} and applying the Dyson's equation between slices $M+1$ and 0 gives
\begin{equation}
\langle\psi_{M+1}|\psi_{\alpha}^{\text{s}}\rangle=\mathcal{G}_{M+1,0}U_{0,1}\Psi_1e^{ik_{\alpha}^+}-\mathcal{G}_{M+1,1}U_{1,0}\Psi_0.
\end{equation}
On the other hand, from \eqref{eq:s}
\begin{equation}
\langle\psi_{M+1}|\psi_{\alpha}^{\text{s}}\rangle=\Psi_1\sum_{\beta}s_{\beta\alpha}.
\end{equation}
The Dyson's equation further gives $\mathcal{G}_{M+1,0}=-\mathcal{G}_{M+1,1}U_{1,0}\Gamma_l$. As a result, the matrix of transmission amplitudes $\textbf{S}$ is\cite{Xu08}
\begin{equation}
\Psi_1\textbf{S}=-\mathcal{G}_{M+1,0}(U_{0,1} \Psi_1 K_1-\Gamma_l^{-1}\Psi_0),
\label{eq:propS}
\end{equation}
where $\textbf{S}$ has the dimension $N\times N_{\text{prop}}$, $N_{\text{prop}}$ is the number of propagating states in the leads, $\Psi_0$ and $\Psi_1$ are wave functions at 0 and 1 slices, $\Gamma_l$ is given by \eqref{eq:Gl}, $K_1$ is the diagonal matrix with elements $K_{1,\alpha\beta}=e^{ik_\alpha^+}\delta_{\alpha\beta}$. The Green's function $\mathcal{G}_{M+1,0}$ connects $M+1$ and 0 slices, which are the ending slices of the leads attached to the scattering region, Fig. \ref{fig:B}(c). To calculate $\mathcal{G}_{M+1,0}$ the standard recursion algorithm is used,\cite{Xu08, Ana08, Mac85, Sol89} which is the more efficient method than direct matrix inversion as Eq. \eqref{eq:Gdefinition} might otherwise suggest.\cite{Datta} 

The matrix of the reflection amplitudes $\textbf{R}$ is derived similarly to \eqref{eq:propS} and reads\cite{Xu08}
\begin{equation}
\Psi_0\textbf{R}=-\mathcal{G}_{0,0}(U_{0,1} \Psi_1 K_1-\Gamma_l^{-1}\Psi_0) - \Psi_0.
\label{eq:propR}
\end{equation}
Together with the transmission amplitudes, the scattering matrix is completely determined and satisfies the unitarity condition implied by current conservation.\cite{Datta} 

The sum over transmission and reflection coefficients gives the number of channels open for propagation in the lead, $M_{\text{lead}}$, the so-called sum rule\cite{Datta}
\begin{equation}
\sum_{\alpha\beta} (|t_{\beta\alpha}|^2+|r_{\beta\alpha}|^2)=M_{\text{lead}}.
\label{eq:sumrule}
\end{equation}

\subsection{Wave functions}

The calculation of the wave function inside the scattering region proceeds in two steps. First, the wave functions in the leads, $|\psi_{\alpha}^\text{i}\rangle+|\psi_{\alpha}^\text{r}\rangle$ and $|\psi_{\alpha}^\text{t}\rangle$, Eqs. \eqref{eq:states}, are determined from the transmission and reflection amplitudes \eqref{eq:propS}, \eqref{eq:propR}. The second step is recursive and requires the Green's functions in the scattering region. It starts from Eq. \eqref{eq:psiM} to obtain $\psi_M$, the wave function at the slice $M$ next to the right lead, Fig. \ref{fig:B}(c). Eq. \eqref{eq:psiM} can be rewritten in a general form 
 \begin{equation}
    \psi_i =\mathcal{G}_{i,1}U_{1,0}\psi_0 + \mathcal{G}_{i,i}U_{i,i+1}\psi_{i+1}, \label{eq:psii}
\end{equation}
and then applied again but for the slice $i=M-1$. By this way the recursion continues backward until the slice $i=1$. Alternatively, one can start from Eq. \eqref{eq:psi1} and recurse forward to obtain the wave functions inside the scattering region.

\subsection{Validity checks}

The validity of the above method and its numerical implementation has been checked on several tests.
The sum rule \eqref{eq:sumrule}, and thus unitarity of the scattering matrix and current conservation,\cite{Datta} is fulfilled with an accuracy greater than $10^{-2}$. At $B=0$ T, the conductance through a single constriction as presented in Ref. \onlinecite{Mun08} was reproduced identically. Similarly, the wave functions and dispersion relation for ideal GNRs, both in armchair and zigzag orientations, were obtained in the quantitative agreement with the previous results.\cite{Nak96, Bre06, Zhe07} At finite $B$, magnetic depopulation\cite{Datta, Bee91} of electron quantization subbands was calculated in agreement with Refs. \onlinecite{Dat08, Gui12, Wur10}. The dispersion relations shown in Fig. \ref{fig:A} are qualitatively similar to the ones presented in Refs. \onlinecite{Bre06-2, Cas09}. Landau levels, which can be traced in GNR bulk in Fig. \ref{fig:A}, agree reasonably well with the analytical result of the Dirac equation for 2D graphene,\cite{Cas09} provided that the energy levels in GNR is a result of both magnetic field and finite-size confinement.

\twocolumngrid


\begin{thebibliography}{99}
\bibitem{largeB} The large absolute values of $B$ used throughout this work is a consequence of relatively narrow geometry, 50 nm vs. few $\mu$m in real devices,\cite{Ron21, Dep21} that requires condition $l_B\ll W$ for QHE; for wider geometries, $B$ would be smaller according to the scaling rules.\cite{Sol89}
\end{thebibliography}
\end{document}